\documentclass[onecolumn, a4paper]{IEEEtran}
\IEEEoverridecommandlockouts

\usepackage{mleftright}       
\mleftright                   

\usepackage{graphicx}         
\usepackage{booktabs} 

\usepackage[utf8]{inputenc}
\usepackage{amsmath,amssymb,amsfonts}
\usepackage{mathtools}
\usepackage{amsthm}
\usepackage{mathrsfs}
\usepackage{algorithmic}
\usepackage{ragged2e}
\usepackage{textcomp}
\usepackage{tikz}
\usepackage{caption}
\usepackage{cuted}
\usepackage{pgfgantt}
\usepackage{pdflscape}
\usepackage{pst-plot}
\usepackage{comment} 
\usepackage{cases}
\usepackage{nccfoots}
\usepackage{lineno}
\usepackage{graphicx}
\usepackage{float}
\usepackage[caption=false,font=footnotesize]{subfig}
\usetikzlibrary{spy}
\usetikzlibrary{positioning,calc}
\usetikzlibrary{decorations.pathmorphing,calc,shapes,shapes.geometric,patterns}
\usetikzlibrary{shapes.multipart}
\usepackage{xfrac}
\usepackage{colortbl}
\usepackage{cancel} 
\usepackage{bbm}
\usetikzlibrary{arrows,positioning,calc,intersections}
\usetikzlibrary{datavisualization.formats.functions}
\def\BibTeX{{\rm B\kern-.05em{\sc i\kern-.025em b}\kern-.08em
    T\kern-.1667em\lower.7ex\hbox{E}\kern-.125emX}}
    
\usepackage{romannum}
\usepackage{pgfplots}
\usepgfplotslibrary{fillbetween}
\usetikzlibrary{arrows, decorations.markings}
\usetikzlibrary{arrows.meta}

\newtheorem{theorem}{Theorem}
\newtheorem*{theorem*}{Theorem}
\newtheorem{lemma}{Lemma}

\newtheorem{definition}{Definition}

\newtheorem{remark}{Remark}

\renewcommand{\baselinestretch}{1}

\DeclarePairedDelimiterX{\infdivx}[2]{(}{)}{%
  #1\;\delimsize\|\;#2%
}

\pgfplotsset{compat=1.18}
\begin{document}
\title{Joint Identification and Sensing with Noisy Feedback: A Task-Oriented Communication Framework for 6G} 



\author{
\IEEEauthorblockN{
Yaning Zhao\IEEEauthorrefmark{1},
Holger Boche\IEEEauthorrefmark{2},
Christian Deppe\IEEEauthorrefmark{1}}
\IEEEauthorblockA{\IEEEauthorrefmark{1}Technical University of Braunschweig,
\IEEEauthorrefmark{2}Technical University of Munich\\
Email: yaning.zhao@tu-braunschweig.de, boche@tum.de, christian.deppe@tu-braunschweig.de\vspace{-2mm}}
}

\maketitle


\begin{abstract}
Task-oriented communication is a key enabler of emerging 6G systems, where the objective is to support decisions and actions rather than full message reconstruction. From an information-theoretic perspective, identification (ID) codes provide a natural abstraction for this paradigm by enabling receivers to test whether a task-relevant message was sent, without decoding the entire message. Motivated by the strong impact of feedback on ID and by the growing interest in integrated communication and sensing, this paper studies joint identification and sensing (JIDAS) over state-dependent discrete memoryless channels with noisy strictly causal feedback. The transmitter conveys identification messages while simultaneously estimating the channel state from the feedback signal. For both deterministic and randomized coding schemes, we derive lower and upper bounds on the capacity--distortion function. The results quantify the fundamental limits of JIDAS under noisy feedback and recover existing noiseless-feedback characterizations as special cases.
\end{abstract}

\vspace{-2mm}
\section{Introduction}
\label{sec:intro}

The mathematical theory of communication, initiated by Shannon \cite{shannon1948mathematical}, established the foundations of reliable message transmission and has guided communication system design for decades. In many emerging 6G applications, however, the objective is not the faithful reconstruction of a transmitted message, but the execution of a task based on communicated information. Examples include autonomous driving, industrial automation, and machine-type coordination, where a receiver often needs to verify, trigger, or update an action rather than decode a high-rate payload \cite{boche2018secure,lu2017industry,fettweis20226g,cabrera20216g}. This shift has motivated growing interest in task-oriented communication.

From an information-theoretic viewpoint, identification (ID) coding provides a natural formalization of task-oriented communication. Introduced by Ahlswede and Dueck \cite{ahlswede1989identification} and preceded by related work by Jaja \cite{ja1985identification}, the ID paradigm replaces full-message decoding by a binary decision test of the form ``Was message \(i\) sent?'' for each message index \(i\). This abstraction is well aligned with task execution and leads to substantially different asymptotic behavior from classical transmission. In particular, randomized ID codes permit a doubly exponential number of identifiable messages, and deterministic ID can also outperform classical transmission rates on discrete memoryless channels (DMCs). Moreover, ID coding interacts in a distinctive way with auxiliary resources such as common randomness and feedback \cite{ahlswede1989identificationfeedback,ezzine2021common,ahlswede1995new,labidi2020secure}.

Feedback is especially important in ID. In contrast to classical message transmission, where even perfect feedback does not increase the capacity of a DMC \cite{wolfowitz2012coding}, even noisy feedback can fundamentally enhance identification performance by enabling adaptive code constructions and by generating common randomness between transmitter and receiver. Most existing information-theoretic studies of ID and joint identification-sensing, however, assume idealized noiseless feedback. This assumption is often unrealistic in practical systems, where feedback links are corrupted by measurement noise, signaling errors, quantization, or environmental disturbances. A key motivation of this paper is therefore to understand how noisy feedback alters the fundamental limits of joint identification and sensing.

In parallel, sensing and environment awareness are becoming integral components of future communication systems \cite{bourdoux20206g,Fettweis2021}. Recent work has shown that communication and sensing should be designed jointly rather than separately, leading to fundamental capacity--distortion tradeoffs for channels with state uncertainty and feedback \cite{kobayashi2018joint,9834554,10153971}. In this context, joint identification and sensing (JIDAS) combines task-oriented communication with state estimation: the transmitter aims to convey an identification message while simultaneously estimating the channel state from feedback observations. This model is particularly relevant in closed-loop systems in which reliable decision support and environment awareness must coexist.

Existing JIDAS results focus on noiseless feedback \cite{zhao2024achievable,labidi2024joint}. In contrast, this paper studies JIDAS over state-dependent DMCs with \emph{noisy strictly causal feedback}. The noisy-feedback setting is not merely a technical extension: the encoder no longer observes past channel outputs directly, but only a corrupted version of them. As a result, both the generation of common randomness and the sensing performance are fundamentally affected. Clarifying these effects is essential for moving JIDAS from an idealized theoretical model toward a practically meaningful task-oriented communication framework.

The main contributions of this paper are as follows:
\begin{enumerate}
    \item We formulate JIDAS over state-dependent DMCs with noisy strictly causal feedback;
    
    \item We establish lower and upper bounds on the capacity--distortion function for both deterministic and randomized identification schemes;
    
    \item We provide a binary example illustrating the capacity--distortion tradeoff and the gain over conventional time-sharing.
\end{enumerate}

\emph{Organization:} Section~\ref{sec:main} introduces the system model and main results. Section~\ref{sec:proof} presents the achievability arguments. The converse is given afterward, followed by a numerical example in Section~\ref{sec:example}. Section~\ref{sec:conclusion} concludes the paper.

\emph{Notation:} For a random variable \(X\), let \(P_X\) denote its probability mass function. We write \(H(\cdot)\), \(\mathbb{E}[\cdot]\), and \(I(\cdot;\cdot)\) for entropy, expectation, and mutual information, respectively. All mutual information quantities are evaluated with respect to the joint distribution induced by the channel and the coding scheme; when needed, this distribution is specified explicitly. The notation \([N]\) stands for the set \(\{1,\ldots,N\}\).

\section{System Model and Main Results}
\label{sec:main}
\begin{figure}[H]
\centering
\hspace{-1.1cm}
    \scalebox{1}{

\tikzstyle{farbverlauf} = [ top color=white, bottom color=white]
\tikzstyle{block} = [draw,top color=white, bottom color=white, rectangle, minimum height=2em, minimum width=3em]
\tikzstyle{block1} = [draw, fill=none, rectangle,minimum height=12em, minimum width=2cm]
\tikzstyle{block2} = [draw, fill=none, rectangle,minimum height=2em, minimum width=2em]
\tikzstyle{input} = [coordinate]
\tikzstyle{sum} = [draw, circle,inner sep=0pt, minimum size=2mm,  thick]

\scalebox{1}{
\tikzstyle{arrow}=[draw,->] 
\begin{tikzpicture}[auto, node distance=2cm,>=latex']
\node[] (M) {$i \in \mathcal{N}$};

\node[block,right=.5cm of M] (enc) {Encoder};
\node[block, above=.7cm of enc] (est) {Estimator};
\node[left=.4cm of est] (S) {$\hat{S}_{t-1}$};
\node[block, right=3.2cm of enc] (channel) {$W(y,z|x,s)$};
\node[block1, dashed] at (1.75,.83) (tr) {};
\node[above=.1cm of tr] (tran) {sender};
\node[block,below=.7cm of channel](state){$P_S$};
\node[block, right=1cm of channel] (dec) {Decoder};
\node[below=.4cm of dec] (j) {$j \in \mathcal{N}$};
\node[right=.4cm of dec] (Mhat) {\begin{tabular}{c} $i=j?$ \end{tabular}};
\node[input,right=.7cm of channel] (t1) {};
\node[block2,above=2cm of channel] (t4) {D};
\node[below=1.5cm of t1] (t5) {};
\node[input,above=2.3cm of channel] (t2) {};
\node[input,left=2cm of t2] (point) {};
\node[input,below=1.8cm of point] (ttpoint) {};
\node[input,left=3.5cm of t2] (tt) {};
\node[input,below=1.8cm of tt] (ttt) {};
\node[input,left=0.4cm of ttt] (ttn) {};
\node[input, right=.5cm of enc] (t3) {};
\node[input, below=0.35 cm of est] (tte) {};
\draw[-{Latex[length=1.5mm, width=1.5mm]},thick] (M) -- (enc);
\draw[-{Latex[length=1.5mm, width=1.5mm]},thick] (enc) --node[above]{ $X_t=f_i^t(Z^{t-1})$} (channel);
\draw[-{Latex[length=1.5mm, width=1.5mm]},thick] (channel) --node[above]{$Y_t$} (dec);
\draw[-{Latex[length=1.5mm, width=1.5mm]},thick] (dec) -- (Mhat);
\draw[-] (channel) -- (t4);
\draw[-{Latex[length=1.5mm, width=1.5mm]},thick] (t4) -| node[left]{$Z_{t-1}$} (est);
\draw[-{Latex[length=1.5mm, width=1.5mm]},thick] (state)--node[left]{$S_t$}(channel);
\draw[-{Latex[length=1.5mm, width=1.5mm]},thick] (est) --(S);
\draw[-{Latex[length=1.5mm, width=1.5mm]},thick] (t3) |- (est);
\draw (ttt) arc[start angle=0, end angle=180, radius=0.2]  (ttn); 
\draw[-{Latex[length=1.5mm, width=1.5mm]},thick] (ttn) -| (enc);
\draw[-] (point) -- (ttpoint);
\draw[-] (ttpoint) -- (ttt);
\draw[-{Latex[length=1.5mm, width=1.5mm]},thick] (j)--(dec);
\end{tikzpicture}}
}
    \caption{System model}
    \label{fig:system_model}
\end{figure}
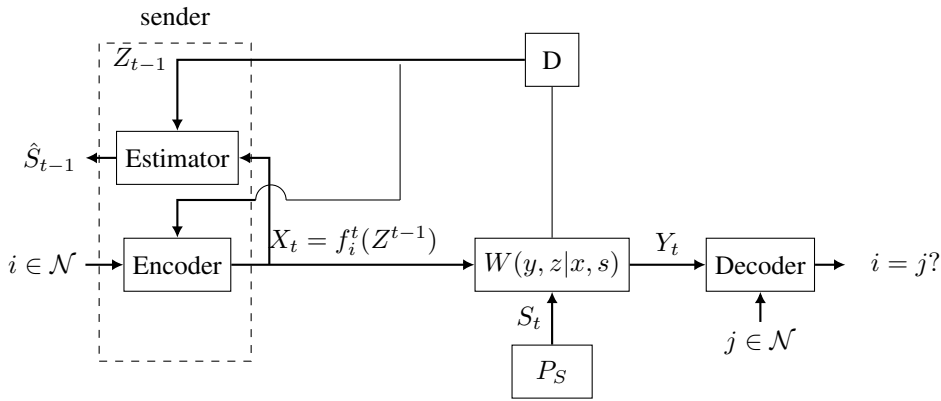

Consider the setup shown in Fig.~\ref{fig:system_model}. At time \(t\), the channel input, decoder output, feedback output, and channel state are denoted by \(X_t\in\mathcal{X}\), \(Y_t\in\mathcal{Y}\), \(Z_t\in\mathcal{Z}\), and \(S_t\in\mathcal{S}\), respectively. The state sequence \(S^n\) is i.i.d. according to \(P_S\). The channel is a state-dependent memoryless channel with transition law
\[
W(y,z|x,s), \qquad (x,s,y,z)\in\mathcal{X}\times\mathcal{S}\times\mathcal{Y}\times\mathcal{Z}.
\]
Equivalently, we may write
\[
W(y,z|x,s)=W_{Y|XS}(y|x,s)P_{Z|Y}(z|y),
\]
which makes explicit that the feedback signal \(Z_t\) is a noisy observation of the channel output \(Y_t\). In particular, noiseless feedback is recovered as the special case \(Z_t=Y_t\).
We further define the averaged channel
\[
\hat{W}(y,z|x)
\triangleq
\sum_{s\in\mathcal{S}} P_S(s) W(y,z|x,s),
\]
and the corresponding forward marginal channel
\[
\tilde{W}_{Y|X}(y|x)
\triangleq
\sum_{z\in\mathcal{Z}} \hat{W}_{YZ|X}(y,z|x).
\]
\vspace{-2mm}
\vspace{-2mm}
\begin{remark}
In this paper, feedback is \emph{strictly causal}: at time \(t\), the encoder observes only \(Z^{t-1}\) when choosing \(X_t\). This contrasts with instantaneous feedback, where the encoder also observes the current feedback symbol. Strictly causal feedback is the more realistic model in the presence of processing and propagation delays.
\end{remark}
An $(n,2^{2^{nR}},\lambda_1,\lambda_2)$ ID code with sensing consists of the following components:
\begin{enumerate}
    \item \textbf{Message set}: $[N]=[1:2^{2^{nR}}]$.
    \item \textbf{Encoder:} For each message \(i \in [N]\), the encoder generates a channel input sequence \(x^n\) using a vector-valued feedback encoding function \(\boldsymbol{f}_i = [f_i^1,\ldots,f_i^n]\). The first component satisfies \(f_i^1 \in \mathcal{X}\), and for each \(t \in [2,n]\), \(f_i^t : \mathcal{Z}^{t-1} \to \mathcal{X}\). Thus, the channel input at time \(t\) depends on the past feedback sequence \(z^{t-1}\) and is given by $x_t = f_i^t(z^{t-1})$.
Let \(\mathcal{F}_n\) denote the set of all length-\(n\) feedback encoding functions;
    \item \textbf{Decoder}: Given $j\in[N]$, the decoder outputs a binary decision answering the question $i=j?$;
    \item \textbf{State estimator}: At time $t$, a state estimator maps each pair of input and feedback symbols $(x_{t-1},z_{t-1})$ to an estimated state $\hat{s_{t-1}}\in\mathcal{S}$. The estimation must satisfy the per-symbol distortion constraint
    \begin{align}\label{eq:distortion}
    \mathbb{E}[d(S_t,\hat{S}_t)]\leq D,\quad \forall t\in[n],
\end{align}
where $d:\mathcal{S}\times \mathcal{S}\mapsto [0,\infty)$ is a distortion function.
Without loss of generality, a deterministic estimator can be used \cite{kobayashi2018joint}, defined as $h:\mathcal{X}\times\mathcal{Z}\mapsto\mathcal{S}$,
\begin{align}
    \hat{s}=h(x,z)=\arg\min_{s'\in\mathcal{S}}\sum_{s\in\mathcal{S}}P_{S|XZ}(s|xz)d(s,s').\nonumber
\end{align}
\end{enumerate}
For each $x\in\mathcal{X}$, define the minimum distortion
\begin{align}
    d^{\star}(x)=\mathbb{E}_{SZ}\left[d\left(S,h\left(x,Z\right)\right)|X=x\right].\nonumber
\end{align}
Similarly, for any input distribution $P_X\in\mathcal{P}(\mathcal{X})$, define
\begin{align}
    d^{\star}(P_X)=\sum_{x\in\mathcal{X}}P_X(x)d^{\star}(x)=\mathbb{E}_{XSZ}\left[d(S,h(X,Z))\right].\nonumber
\end{align}



We next define the deterministic identification feedback (DIF) and randomized identification feedback (RIF) codes, and present the main results for JIDAS over state-dependent DMCs.
\begin{definition}
    Let $\lambda_1,\lambda_2\geq 0$ with $\lambda_1+\lambda_2<1$. An $\left(n,N_F,\lambda_1,\lambda_2\right)$ DIF code for the channel $W(y,z|x,s)$ is a family $\left\{\left(\boldsymbol{f}_i,\mathcal{D}_i\right)|i\in[N]\right\}$ with feedback encoding functions $\boldsymbol{f}_i\in\mathcal{F}_n$ and decoding regions $\mathcal{D}_i\subset \mathcal{Y}^n$; such that for all $i\in [N]$ and $i\neq \tilde{i}\in[N]$, the Type I error $P_{e,1}(i)$ and the Type II error $P_{e,2}(i,\tilde{i})$ are bounded as follows:
    \begin{align}
        \sum_{z^n\in\mathcal{Z}^n}\sum_{s^n\in\mathcal{S}^n}P^n_S(s^n)W^n(\mathcal{D}_i^c,z^n|\boldsymbol{f}_i,s^n)&\leq \lambda_1,\nonumber\\
        \sum_{z^n\in\mathcal{Z}^n}\sum_{s^n\in\mathcal{S}^n}P^n_S(s^n)W^n(\mathcal{D}_{\tilde{i}},z^n|\boldsymbol{f}_i,s^n)&\leq \lambda_2.\nonumber
    \end{align}
\end{definition}

Each code can be characterized by the double-exponential rate scale as $R=\frac{\log{\log{N}}}{n}$. A rate is said to be achievable, if there exists an $(n,N,\lambda_1,\lambda_2)$ DIF code that can attain it. Let $N_{DIF}(n,\lambda_1,\lambda_2,D)$ be the maximum integer for which a $(n,N_{DIF},\lambda_1,\lambda_2)$ DIF code exists and satisfies the distortion constraint in \eqref{eq:distortion}. The DIF capacity-distortion function $C_{DIF}(D)$ is defined as 
\begin{align}
    C_{DIF}(D)=\inf_{\substack{\lambda_1,\lambda_2>0\\\lambda_1+\lambda_2<1}}\liminf_{n\to\infty}\frac{\log\log{N_{DIF}}(n,\lambda_1,\lambda_2,D)}{n}.\nonumber
\end{align}
\begin{theorem} \label{thm:JDIFAS}
If the transmission capacity of the marginal channel \(\tilde{W}_{Y|X}\) is positive, then for all \(\lambda_1,\lambda_2 \ge 0\) with \(\lambda_1+\lambda_2<1\),
\begin{align}
    C_{DIF}(D)\geq 
    \max_{x\in\mathcal{X}_D,P_{U|X=x,Z}:I(U;Z|X,Y)<I(X;Y)} I(U;Z|X=x)\nonumber,
\end{align}
 and
\begin{align}
    C_{DIF}(D)
    &\leq \min\left\{\max_{x\in\mathcal{X}_D} I (Y;Z|X=x),\max_{x\in\mathcal{X}_D} H(Z|X=x)\right\},\nonumber
\end{align}
where $\mathcal{X}_D \triangleq \{x\in\mathcal{X}: d^{\star}(x)\le D\}$. 

If the transmission capacity of \(\tilde{W}_{Y|X}\) is zero, then \(C_{DIF}(D)=0\).
\end{theorem}
\begin{definition}
    Let $\lambda_1,\lambda_2\geq 0$ and $\lambda_1+\lambda_2<1$. Then, an $\left(n,N_R,\lambda_1,\lambda_2\right)$ RIF code for channel $W(y,z|x,s)$ is a family $\left\{\left(Q_F(\cdot|i),\mathcal{D}_i\right)|i\in[N]\right\}$ with $Q_F(\cdot|i)\in\mathcal{P}(\mathcal{F}_n)$ and decoding regions $\mathcal{D}_i\subset \mathcal{Y}^n$; such that the Type I error $P_{e,1}(i)$ and the Type II error $P_{e,2}(i,\tilde{i})$ are bounded as follows:
    \begin{align}
        \sum_{\boldsymbol{f}\in\mathcal{F}_n}Q_F(\boldsymbol{f}|i)\sum_{z^n\in\mathcal{Z}^n}\sum_{s^n\in\mathcal{S}^n}P^n_S(s^n)W(\mathcal{D}_i^c,z^n|\boldsymbol{f},s^n)&\leq \lambda_1,\nonumber\\
        \sum_{\boldsymbol{f}\in\mathcal{F}_n}Q_F(\boldsymbol{f}|i)\sum_{z^n\in\mathcal{Z}^n}\sum_{s^n\in\mathcal{S}^n}P^n_S(s^n)W(\mathcal{D}_{\tilde{i}},z^n|\boldsymbol{f},s^n)&\leq \lambda_2.\nonumber
    \end{align}
\end{definition}
Similarly, the RIF capacity-distortion function $C_{RIF}(D)$ is defined as
\begin{align}
    C_{RIF}(D)=\inf_{\substack{\lambda_1,\lambda_2>0\\\lambda_1+\lambda_2<1}}\liminf_{n\to\infty}\frac{\log\log{N_{RIF}}}{n},\nonumber
\end{align}
where $N_{RIF}(n,\lambda_1,\lambda_2,D)$ is the maximum number of messages in the code. In this work we prove the following:
\begin{theorem}\label{thm:JRIDAS}
    If the transmission capacity of the forward marginal channel \(\tilde{W}_{Y|X}\) is positive, then for all \(\lambda_1,\lambda_2 \ge 0\) with \(\lambda_1+\lambda_2<1\),
\begin{align}
    \max_{P_X\in\mathcal{P}_D,P_{U|X,Z}: I(U;Z|X,Y)<I(X;Y)} I(X,U;Y)\leq C_{RIF}(D)\nonumber\\
    \leq \max_{P_X\in\mathcal{P}_D} I ((X,Z);Y),\nonumber
\end{align}
where $\mathcal{P}_D\triangleq \left\{P\in\mathcal{P}(\mathcal{X}):d^{\star}(P)\leq D\right\}$ and the mutual information terms are evaluated with respect to the joint distribution $P_{XYZ}(x,y,z)=P_X(x)\hat{W}_{YZ|X}(y,z|x)$.

If the transmission capacity of \(\tilde{W}_{Y|X}\) is zero, then \(C_{RIF}(D)=0\).
\end{theorem}

\begin{remark}
When the feedback is noiseless, i.e., \(Z=Y\), the bounds in Theorems~\ref{thm:JDIFAS} and \ref{thm:JRIDAS} reduce to the corresponding results for JIDAS with noiseless feedback in \cite{labidi2024joint}. This shows that feedback noise reduces the amount of useful common randomness available for identification and sensing.
\end{remark}
\begin{remark}
If the sensing requirement is removed, the model reduces to identification with feedback, and the distortion constraint is eliminated. Consequently, the achievable identification rate is generally higher.
\end{remark}
From a system perspective, JIDAS captures a perception--communication--decision loop in which noisy feedback supports state estimation at the transmitter, while identification enables decision-oriented signaling. This abstraction is relevant for closed-loop 6G systems in which communication is designed to support control, coordination, and inference rather than raw data delivery \cite{kutsevol2023goal,perotti2024identification}.
\section{Coding Scheme for the Achievability}
\label{sec:proof}
We extend the coding scheme for JIDAS with noiseless feedback over DMCs proposed in \cite{labidi2024joint} to the case with noisy feedback. In the original scheme, the first $n$ channel uses are employed to establish uniform common randomness between the transmitter and receiver through the feedback link, followed by $\lceil\sqrt{n}\rceil$ channel uses for message identification. In our setting, we show that even with noisy feedback, it remains possible to identify a doubly exponential number of messages reliably while satisfying the prescribed distortion constraint.

\subsection{Achievability proof of Theorem \ref{thm:JDIFAS}}
\label{subsec:DIFdirect}
\emph{Code construction:} We begin with an $\left(m,N,\lambda_1,\lambda_2\right)$ DIF code $\left\{(\boldsymbol{f}_i,\mathcal{D}_i):i\in[N]\right\}$, where $m=n+\lceil\sqrt{n}\rceil$. The first $n$ channel uses are based on a superposition coding structure:
\begin{enumerate}
     \item Base code $\mathcal{C}_b$: Define $\mathcal{C}_b$ as an $\left(n,M_b=2^{n(I(X;Y)+\epsilon)},2^{-n\delta_b}\right)$ transmission code $\mathcal{C}_b\subset\mathcal{X}_D^n$.
Partition $\mathcal{C}_b$ into $L=2^{n(I(U;Z|X,Y)+\gamma)}$ subcodes, each of size $B=2^{n(I(X;Y)-I(U;Z|X,Y)-\epsilon-\gamma)}$.
    \item Satellite code $\mathcal{C}_s(\boldsymbol{c})$: For each cloud center $\boldsymbol{c}\in\mathcal{C}_b$, generate a satellite code $\mathcal{C}_s(\boldsymbol{c})\subset\mathcal{U}^n$ of $\left(n,M_s,2^{-n\delta_s}\right)$, where each codeword $u^n$ is drawn according to $P_{U|X=x}$. For fixed $\boldsymbol{c}\in\mathcal{C}_b$ and $z^n\in\mathcal{Z}^n$, define the subset $\mathcal{T}_c(z^n)=\mathcal{T}_{\epsilon}^n(P_{ZU}|z^n)$ with the mapping $T_{\boldsymbol{c}}:\mathcal{Z}^n\mapsto\mathcal{T}_{\boldsymbol{c}}$. 
    \item Codebook compression: For each $\boldsymbol{c}\in\mathcal{C}_b$, compress $\mathcal{C}_s(\boldsymbol{c})$ into an index set via an encoding function $\Phi_c: \mathcal{C}_s(\boldsymbol{c})\mapsto \left\{1,\cdots,L\right\}$ and decoding function $\Psi_c: \mathcal{C}_b\times\mathcal{Y}^n\times \left\{1,\cdots,L\right\}\mapsto \mathcal{C}_s(\boldsymbol{c})$. By the Slepian–Wolf theorem, the rate $\frac{1}{n}H(T_{\boldsymbol{c}}(Z^n)|cY^n)$ is achievable, enabling recovery of $u^n$ with $\text{Pr}\left(\Psi_c(b,y^n)\neq u^n|\boldsymbol{c}\right)\leq 2^{-n\delta}$. 
    \end{enumerate}
\emph{Common randomness generation:} The encoder first transmits $n$ repetitions of the symbol $x^{\star}=\arg\max_{x\in\mathcal{X}_D} I (U;Z|X=x)$. Let $\boldsymbol{c}^{\star}={x^{\star}}^n$ denote the corresponding base codeword, and $l^{\star}$ its subcode index. Given the feedback sequence $z^n$, the encoder finds a codeword ${u^{\star}}^n\in \mathcal{T}_{\boldsymbol{c}\star}(z^n)$, which serves as the common randomness shared between the encoder (original $u^n$) and the decoder (through $\hat{u}^n=\Psi_{c^{\star}}(l^{\star},y^n)$). The RV $U^n$ is asymptotically uniformly distributed over $\mathcal{T}_{c\star}(Z^n)$, as stated below.
\begin{lemma}For a deterministic codeword $\boldsymbol{c}^{\star}={x^{\star}}^n$, a realization of the noisy feedback sequence $Z^n$, and a typical sequence $u^n\in \mathcal{T}_{\boldsymbol{c}^{\star}}(Z^n)$, we have
    \begin{align}
        P^n_{U}\left(u^n\right)&\doteq 2^{-nI(U;Z|X=x^{\star})},\nonumber\\
        |\mathcal{T}_{\boldsymbol{c}^\star}(Z^n)|&\doteq 2^{nI(U;Z|X=x^{\star})},\nonumber\\
        \text{Pr}\left[U^n\in \mathcal{T}_{\boldsymbol{c}^\star}(Z^n)\right]&\doteq 1,\nonumber
    \end{align}
    where $\doteq$ denotes the asymptotic equivalence in $n$.
    \end{lemma}
\emph{Message identification:} Now, let $\left\{F_{i}:\mathcal{T}_{\boldsymbol{c}^{\star}} \mapsto [M']|i\in[N]\right\}$ be a family of hashing functions, where each $F_i$ maps $M<N$ sequences in $\mathcal{T}_{\boldsymbol{c}^{\star}}$ uniformly at random to a hash value $w\in[M']$, i.e., $Pr\left[F_i(u^n)=w\right]=\frac{1}{M'}$. The functions $F_{i}$, and in particular their supports, are known in advance to both the encoder and the decoder.

If the message $i\in[N]$ is to be sent, the transmitter computes $F_{i}(u^n)$ and uses the resulting hash $w$ as the seed of an
$\left(\lceil\sqrt{n}\rceil,{M'},2^{-\lceil\sqrt{n}\rceil\delta'}\right)$
transmission code. The corresponding code is $\mathcal{C}'=\left\{\left(\boldsymbol{c}'({w})\in\mathcal{X}_D^{\lceil\sqrt{n}\rceil},\mathcal{D}'_{w}\right)\big|{w}\in[M']\right\}$. At the receiver, to test whether the transmitted message is $j$, the decoder computes $F_{j}(\Psi_{\boldsymbol{c}^{\star}}(b^{\star},y^n))$ and compares it with the decoded hash $\hat{w}$. If they match, i.e., $\hat{w}=F_j(\cdot)$, the decoder accepts the hypothesis that message $j$ was sent.

\emph{Error analysis:} The Type I and Type II error probabilities for all $i\neq\tilde{i}\in[N]$ can be bounded as follows.
\begin{align}
    P_{e,1}(i)
    &=\sum_{z^n\in\mathcal{Z}^m}\sum_{s^m\in\mathcal{S}^m}P^m_S(s^m)W^m(\mathcal{D}_i^c,z^n\big|\boldsymbol{f}_i,s^m)\nonumber\\
    &\leq 2^{-n\delta_b}+2^{-n\delta'}=\circ(n)\nonumber.
\end{align}
\begin{align}
    P_{e,2}(i,\tilde{i})
    &= \sum_{z^m \in \mathcal{Z}^m} \sum_{s^m \in \mathcal{S}^m} P_S^m(s^m) 
    W^m\big(\mathcal{D}_{\tilde{i}}, z^m \mid \boldsymbol{f}_i,s^m \big) \nonumber\\
    &\leq 2^{-n\delta_b}+2^{-n\delta'}+\frac{
|\left\{u^n\in\mathcal{T}_{\boldsymbol{c}^{\star}}(Z^n):F_i(u^n)=F_{\tilde{i}}(u^n)\right\}|}{|\mathcal{T}_{\boldsymbol{c}^{\star}}(Z^n)|}.\nonumber
\end{align}
Define an auxiliary random variable $\Gamma_{u^n}$ for each $u^n\in \mathcal{T}_{\boldsymbol{c}^{\star}}(Z^n)$, where
\begin{align*}
    \Gamma_{u^n}(F_{\tilde{i}})=\left\{
        \begin{array}{cc}
             1,&y^n\in F_{i}\cap F_{\tilde{i}}  \\
             0,&y^n\in F_{i}-F_{\tilde{i}}
        \end{array},
    \right.
\end{align*}
with probability $\text{Pr}[\Gamma_{u^n}(F_{\tilde{i}})=1]=\frac{1}{M'}$.

\begin{lemma}\cite{ahlswede1989identificationfeedback}
    For $\lambda \in(0,1)$, and $E[\Psi_{u^n}]=\frac{1}{M'}<\lambda$,
    \begin{align*}
        Pr\left[\frac{1}{|\mathcal{T}_{\boldsymbol{c}^{\star}}(Z^n)|}\sum_{u^n\in \mathcal{T}_{\boldsymbol{c}^{\star}}(Z^n)}\Gamma_{u^n}(F_{\tilde{i}})>\lambda\right]
        <2^{-|\mathcal{T}_{\boldsymbol{c}^{\star}}(Z^n)|\cdot (\lambda\log(M')-1)}.
    \end{align*}
\end{lemma}
For all pairs $(i,\tilde{i})$, in order to achieve an arbitrarily small type-II error probability, it is necessary to upper-bound \\$\frac{|\left\{u^n\in\mathcal{T}_{\boldsymbol{c}^{\star}}(Z^n):F_i(u^n)=F_{\tilde{i}}(u^n)\right\}|}{|\mathcal{T}_{\boldsymbol{c}^{\star}}(Z^n)|}$ by $\lambda_2$. Thus, we obtain the following probability bound:
\begin{align*}
    &\text{Pr}\left[ \bigcap_{\tilde{i}\in\mathcal{N},\tilde{i}\ne i}\left\{\frac{1}{|\mathcal{T}_{\boldsymbol{c}^{\star}}(Z^n)|}\sum_{u^n\in \mathcal{T}_{\boldsymbol{c}^{\star}}(Z^n)}\Gamma_{u^n}(F_{\tilde{i}})\le\lambda\right\}\right]\\
    &\quad \ge 1- (N-1)\cdot 2^{-2^{nI(U;Z|x=x^{\star})}\cdot(\lambda\log{M'}-1)}.
\end{align*}
To ensure that this probability is strictly positive, the number of codewords $N$ must satisfy
\begin{align*}
    N=2^{2^{nI(U;Z|X=x^{\star})}\cdot(\lambda\log{M'}-1)}. 
\end{align*}
This implies that as the code length $m$ goes to infinity, and for sufficiently small $\epsilon$ and $\gamma$, we obtain:
\begin{align*}
    \liminf_{m\to\infty}\frac{\log\log{N_{DIF}}}{m}
    &\geq \frac{\log\log{N}}{n} \nonumber\\
    &= I(U;Z|X=x^{\star}).
\end{align*}
This completes the achievability proof of Theorem~\ref{thm:JDIFAS}.

\subsection{Achievability proof of Theorem \ref{thm:JRIDAS}}
\begin{figure}[H]
    \centering
    \scalebox{0.8}{
\begin{tikzpicture}[>=stealth, thick, scale=1]

\definecolor{orange}{RGB}{255,140,0}
\definecolor{blue}{RGB}{0,90,200}
\definecolor{purple}{RGB}{150,0,200}

\node[draw, minimum width=2cm, minimum height=8cm] (mainbox) {};

\foreach \y in {8,6,4,2}{
  \draw ($(mainbox.south west)!0!(mainbox.south east)+(0,\y)$) -- ++(2cm,0cm);
}
 \draw[decorate,decoration={brace,amplitude=10pt}]
    ($(mainbox.south west)+(0,6)$) -- ($(mainbox.south west)+(0,8)$)
    node[midway,xshift=-0.6cm]{$B$};
 \draw[decorate,decoration={brace,amplitude=10pt}]
    ($(mainbox.south west)+(0,4)$) -- ($(mainbox.south west)+(0,6)$)
    node[midway,xshift=-0.6cm]{$B$};
 \draw[decorate,decoration={brace,amplitude=10pt}]
    ($(mainbox.south west)+(0,2)$) -- ($(mainbox.south west)+(0,4)$)
    node[midway,xshift=-0.6cm]{$B$};
 \draw[decorate,decoration={brace,amplitude=10pt}]
    ($(mainbox.south west)+(0,0)$) -- ($(mainbox.south west)+(0,2)$)
    node[midway,xshift=-0.6cm]{$B$};
    
\node[right=2cm] at ($(mainbox.south west)+(0,5)$) {$\cdots$};
\node[right=2cm] at ($(mainbox.south west)+(0,3)$) {$C_b^{l_{k+1}=\Phi_{c_k}(T_{c_k}(z_k^n))}$};
\node[right=2cm] at ($(mainbox.south west)+(0,1)$) {$C_b^L$};

\node[draw=orange, very thick, minimum width=2cm, minimum height=0.5cm, anchor=north] (ck) at ([yshift=-1cm]mainbox.north) {\color{orange}{$c_k$}};
\node[above=1cm of ck, black] {$C_b \sim P_X\in\mathcal{P}_D$};

\coordinate (rightzoomstart) at ([xshift=1.8cm,yshift=2.5cm]ck.east);
\coordinate (rightzoombottom) at ([xshift=1.8cm,yshift=-2.5cm]ck.east);
\draw[orange, thick] 
  (ck.east) -- (rightzoomstart); 
\draw[orange,thick] ($(rightzoombottom)$)--(ck.east);
\node[draw, minimum width=2cm, minimum height=5cm] at ([xshift=2.8cm,yshift=0cm]ck.east)(subbox) {};
\node[draw, minimum width=2cm, minimum height=2cm,fill=blue!20] at ([xshift=2.8cm,yshift=0cm]ck.east)(code2) {};
\node[draw, blue, thick,minimum width=2cm,minimum height =0.5cm] at ([xshift=2.8cm,yshift=0cm]ck.east){\small $u_k^n=T_{c_k}(z_k^n)$};

\node[above=0cm of subbox, black] {$C_s \sim P_{U|X}$};
\coordinate (p1) at ([xshift=3.8cm,yshift=0cm]ck.east);
\coordinate (p2) at ([xshift=4.5cm,yshift=0cm]ck.east);
\coordinate (p3) at ([xshift=4.5cm,yshift=-4cm]ck.east);
\coordinate (p4) at ([xshift=2.8cm,yshift=-4cm]ck.east);
\draw[purple, thick] 
  (p1) -- (p2) -- (p3) edge[->, purple, thick] (p4); 
\node[draw, minimum width=2cm, minimum height=2cm,fill=purple!15] (code1) at ([yshift=-3.5cm]ck.south){};
\node[draw=purple, very thick, minimum width=2cm, minimum height=0.5cm, anchor=north] (ck1) at ([yshift=-3.5cm]ck.south) {\color{purple}{$c_{k+1}$}};

\end{tikzpicture}
}
    \caption{Superposition coding structure}
    \label{fig:code}
\end{figure}
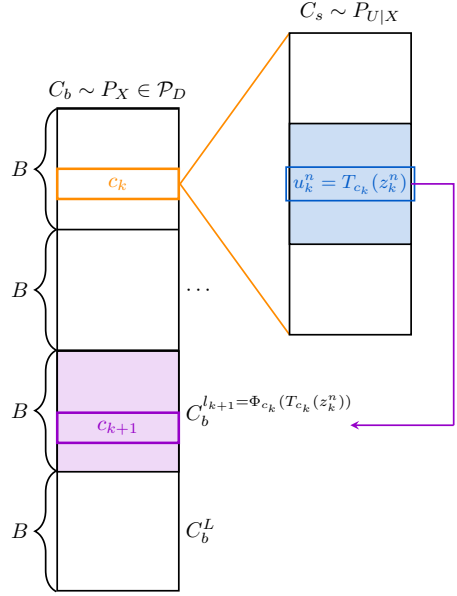

\emph{Code construction:} We start with an $(m,N,\lambda_1,\lambda_2)$ RIF code $\left\{(\boldsymbol{f}_i,\mathcal{D}_i):i\in[N]\right\}$, where $m=nK+\lceil\sqrt{n}\rceil$. The construction follows the superposition coding scheme introduced in Section~\ref{subsec:DIFdirect}, with the following modification (see Fig.~\ref{fig:code}): the satellite codeword $u^n$ is generated according to $\sum_{x\in\mathcal{X}}P_XP_{U|X}$, where $P_X\in\mathcal{P}_D$. We define $\mathcal{T}_{\boldsymbol{c}}(Z^n)\triangleq\mathcal{T}_{\epsilon}(P_{XZU}|x^n)$.

\emph{Common randomness generation and message identification:} The encoder proceeds as follows:
\begin{enumerate}
    \item Transmit  a fixed codeword $\boldsymbol{c}_1\in\mathcal{C}_b$.
    \item Given the feedback sequence $z_1^n$, randomly and uniformly select a codeword $u_1^n=T_{c_1}(z_1^n)\in \mathcal{T}_{\boldsymbol{c}_1}(z_1^n)$. Compute $l_2=\Phi_{\boldsymbol{c}_1}(f_{c_1}(z_1^n))$ and select $\boldsymbol{c}_2$ randomly and uniformly from the subcode $\mathcal{C}^{l_2}_b$. Denote its index by $b_2$;
    \item For each $k\in[K]$, given $z^n_{k-1}$, encode $u^n_k=T_{\boldsymbol{c}_{k}}(z^n_{k})$. Then choose $\boldsymbol{c}_k$ uniformly from the subcode $\mathcal{C}_1^{l_k=\Phi_{\boldsymbol{c}_{k-1}}(T_{\boldsymbol{c}_{k-1}}(z_{k-1}^n))}$ and denote its index by $b_k$.
    \item Finally, Compute $F_i(T_{\boldsymbol{c}_1}(Z_1^n),\cdots,T_{\boldsymbol{c}_{K-1}}(Z_{K-1}^n),$ $b_2,\cdots,b_K)$, encode $F_i$ using $\mathcal{C}_b$, and transmit it through the channel.
\end{enumerate}
\vspace{-2mm}
\vspace{-2mm}
We introduce the following lemma.
\begin{lemma}For a random-generated code word $\boldsymbol{c}=x^n\in\mathcal{X}^n$, a realization of the noisy feedback sequence $Z^n$, and a typical sequence $u^n\in \mathcal{T}_{\boldsymbol{c}}(Z^n)$, we have
    \begin{align}
        P^n_{U}\left(u^n\right)&\doteq 2^{-nI(U;Z|X)},\nonumber\\
        |\mathcal{T}_{\boldsymbol{c}}(Z^n)|&\doteq 2^{nI(U;Z|X)},\nonumber\\
        \text{Pr}\left[U^n\in \mathcal{T}_{\boldsymbol{c}}(Z^n)\right]&\doteq 1,\nonumber
    \end{align}
    where $\doteq$ denotes asymptotic equivalence as $n\to \infty$.
    \end{lemma}
Denote the common randomness sequences $(u_1^n,\dots,u_{K-1}^n,b_2,\cdots,b_K)$ as a vector-valued RV $\boldsymbol{\Theta}$, uniformly distributed over $\vartheta=\mathcal{T}_{\boldsymbol{c}_1}(Z_1^n)\times\cdots\times\mathcal{T}_{\boldsymbol{c}_{K-1}}(Z^n_{K-1})\times \left\{1,\cdots B\right\}^{K-1}$ with $\text{Pr}(\boldsymbol{\Theta}=\boldsymbol{\theta})=2^{-n(I(U;Z|X)+I(X;Y)-I(U;Z|X,Y)-\epsilon-\gamma)}$, for all $\theta\in\vartheta$.
The decoder reconstructs $\left\{\hat{c_1},\hat{c}_2,\cdots,\hat{c}_K\right\}$ and indices $\left\{\hat{b}_2,\cdots,\hat{b}_K\right\}$.
Given $\hat{c}_k$ $y_k^n$ and $\hat{b}_k$, it recovers $\hat{u}_k^n=\Psi_{\hat{c}}(y^n_k,\hat{b}_k)$ and forms $\hat{\boldsymbol{\Theta}}=\left\{\hat{u}_1^n,\hat{u}_2^n,\cdots,\hat{u}_{K-1}\right\}$. It then computes $F_{j}(\hat{\boldsymbol{\theta}})$ and decodes $\hat{F}_i$ from $y^{(n+1)K}_{nK+1}$; if $\hat{F}_i=F_j$, decide $i=j$.

\emph{Error analysis:} For all \(i \in [N]\), it can be shown that that \(P_{e,1}(i) \le \lambda_1\). For any \(i \neq \tilde{i} \in [N]\), the type-II error probability is bounded as follows:
\begin{align}
    P_{e,2}(i,\tilde{i})
    &\leq\frac{|\left\{\boldsymbol{\theta}\in\vartheta:F_i(\boldsymbol{\theta})=F_{\tilde{i}}(\boldsymbol{\theta})\right\}|}{|\vartheta|}+\circ(n).\nonumber
\end{align}
Then
\begin{align*}
    &\text{Pr}\left[ \bigcap_{\tilde{i}\in\mathcal{N},\tilde{i}\ne i}\left\{\frac{1}{|\vartheta|}\sum_{\boldsymbol{\theta}\in \vartheta}\Gamma_{\theta}(F_{\tilde{i}})\le\lambda\right\}\right]\\
    & \ge 1- (N-1)\cdot \nonumber\\
    &\quad2^{-2^{n(K-1)(I(U;Z|X)+I(X;Y)-I(U;Z|X,Y)-\epsilon-\gamma)}\cdot(\lambda\log{M'}-1)}.
\end{align*}
Ensuring positivity yields
\begin{align*}
    N=2^{2^{n(K-1)(I(U;Z|X)+I(X;Y)-I(U;Z|X,Y)-\epsilon-\gamma)}\cdot(\lambda\log{M'}-1)}. 
\end{align*}
With small $\epsilon$ and $\gamma$, 
\begin{align*}
    \liminf_{m\to\infty}\frac{\log\log{N_{RIF}}}{m}
    &\geq \lim_{m\to\infty}\frac{\log\log N}{n}\nonumber\\
    &=I(U;Z|X)+I(X;Y)-I(U;Z|X,Y)\nonumber\\
    &\overset{(a)}{=}I(X,U;Y)\nonumber,
\end{align*}
where $(a)$ follows from that $U-(X,Z)-Y$ forms a Markov chain. 

This completes the direct proof of Theorem \ref{thm:JRIDAS}.

\section{Converse Proof}
\label{sec:converse proof}

The converse proof follows the classical identification framework with feedback, adapted to incorporate the distortion constraint via type-class arguments. Throughout the converse, all mutual information quantities are evaluated with respect to the joint distribution induced by \(P_X\) and the averaged channel \(\hat{W}_{YZ|X}\). The converse proof of Theorem~\ref{thm:JDIFAS} is a special case of Theorem~\ref{thm:JRIDAS} and follows directly. We briefly outline the main steps.

\begin{definition}
A probability distribution \( Q \) on \( \mathcal{X} \times \mathcal{Z} \) is called an \( n \)-type if for all \( (x,z) \in \mathcal{X} \times \mathcal{Z} \), 
\( Q(x,z) \in \{\frac{1}{n}, \ldots, \frac{n}{n}\} \).
\end{definition}

\begin{definition}
Let 
\[
\mathcal{S}_f = \{ (x^n,z^n): f(z^n) = x^n,\; d^{\star}(x) < D \}.
\]
For \( i \in [N] \), the probability of a pair \( (x^n,z^n) \in \mathcal{S}_f \) is
\begin{align}
P_i(x^n,z^n)
=\sum_{\boldsymbol{f}\in\mathcal{F}_n} Q(\boldsymbol{f}|i)
\sum_{y^n,s^n} P_S^n(s^n) W^n(y^n,z^n|x^n,s^n). \nonumber
\end{align}
An \( n \)-type \( P \) on \( \mathcal{X}\times\mathcal{Z} \) is \(\epsilon\)-typical if for any \( (x,z) \),
\[
\left|\frac{P(x,z)}{\sum_{z'} P(x,z')} - \sum_{y,s} P_S(s) W(y,z|x,s)\right| \le \epsilon.
\]
Denote the set of all such types by \( \mathcal{P}^n_\epsilon \).
\end{definition}

The key technical ingredients are two lemmas adapted from \cite{han2002new}:

\begin{lemma}\label{lemma:homogenious}
A homogeneous \( M \)-regular \( (n,N,\lambda_1,\lambda_2) \)-ID code with \( \lambda_1+\lambda_2<1 \) satisfies
\[
\log N \le n(n+1)^{|\mathcal{X}|} M \log |\mathcal{X}|.
\]
\end{lemma}

\begin{lemma}
For any \( (n,N,\lambda_1,\lambda_2) \) feedback ID code with per-symbol distortion constraint \( \mathbb{E}[d(S_t,\hat{S}_t)] < D \), and for any \( \lambda'_1>\lambda_1 \), \( \lambda'_2>\lambda_2 \) with \( \lambda'_1+\lambda'_2<1 \), there exists a homogeneous \( e^{n(T+\gamma)} \)-regular \( (n,N',\lambda'_1,\lambda'_2) \)-ID code such that 
\[
N' > N e^{-\delta n(n+1)^{|\mathcal{X}||\mathcal{Z}|}},\quad 
T = \max_{P_X \in \mathcal{P}_D} I(X,Z;Y).
\]
\end{lemma}

Combining both lemmas gives
\[
\log N - \delta n (n+1)^{|\mathcal{X}||\mathcal{Z}|}
< n (n+1)^{|\mathcal{X}|} e^{n(T+\gamma)} \log |\mathcal{X}|.
\]
Hence,
\begin{align}
\lim_{n\to\infty} \frac{\log \log N}{n}
&\le T + \gamma.\nonumber
\end{align}
Since \( \gamma \) is arbitrary, the capacity–distortion function is upper bounded by
\[
C_{\mathrm{RIF}}(D) \le \max_{P_X \in \mathcal{P}_D} I(X,Z;Y).
\]
This completes the converse proof.

\section{Example}
\label{sec:example}
\begin{figure}[H]
    \centering
    \includegraphics[width=0.6\linewidth]{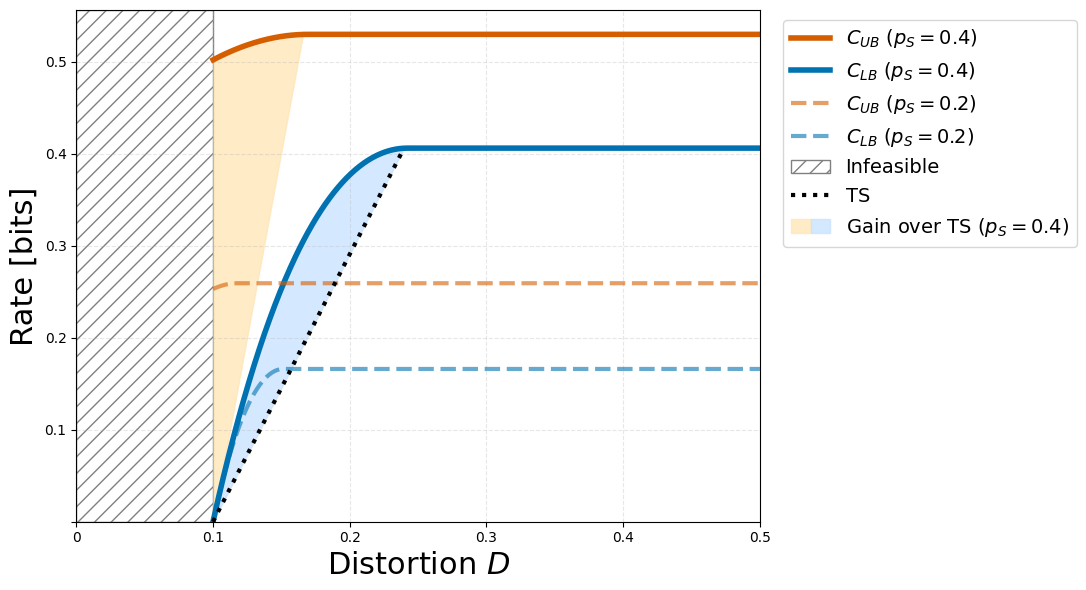}
    \caption{Lower and upper bound of $C(D)$ ($p_N=0.1$)}
    \label{fig:C_D_example}
\end{figure}

We consider a binary channel with a multiplicative Bernoulli state, characterized by $Y=X \cdot S$ and $Z=Y\oplus N$, where $\mathcal{X} = \mathcal{S} = \mathcal{Y} = \mathcal{Z} = \{0,1\}$. The RVs $S$ and $N$ are independent, with $S \sim \mathrm{Bernoulli}(p_S)$, and $N \sim \mathrm{Bernoulli}(p_N)$. We adopt the Hamming distortion measure $d(s,\hat{s}) = s \oplus \hat{s}$ and focus on the parameter regime $ 0 \leq p_N \leq p_S \leq \tfrac{1}{2}$. The upper and lower bounds of the RIF capacity--distortion function, specialized from Theorem~\ref{thm:JRIDAS}, are illustrated in Fig.~\ref{fig:C_D_example} for $p_S=0.2, 0.4$ and $p_N=0.1$, together with the performance of conventional time-sharing (TS) schemes. A clear gain of the proposed co-design scheme over TS is observed. Moreover, the gap between the bounds decreases with increasing $p_S$ and increases with larger feedback noise $p_N$.

\vspace{-2mm}
\vspace{-2mm}

\section{Conclusion}
\label{sec:conclusion}

We studied joint identification and sensing over state-dependent DMCs with noisy strictly causal feedback. For both deterministic and randomized identification with feedback, we derived lower and upper bounds on the capacity--distortion function and clarified how noisy feedback affects the identification--sensing tradeoff.

Several directions remain open. An important next step is to identify conditions under which the lower and upper bounds coincide. It is also of interest to study other non-ideal feedback models, such as rate-limited or delayed feedback, and to extend the analysis from per-symbol to average distortion constraints. Beyond discrete memoryless models, extensions to Gaussian channels, multiuser settings, and quantum communication systems appear particularly promising.

\section*{Acknowledgments}
H. Boche, C. Deppe, and Y. Zhao acknowledge financial support from the Federal Ministry of Research, Technology and Space of Germany (BMFTR) under the program “Souverän. Digital. Vernetzt.”, within the research program Communication Systems Souverän. Digital. Vernetzt., Joint Project 6G-life (project identification numbers 16KIS2414 and 16KIS2415). H. Boche and C. Deppe further acknowledge partial support from the BMFTR through the Q-STARS program (Grants 16KIS2611 and 16KIS2602).
In addition, C. Deppe and Y. Zhao were supported by the German Research Foundation (DFG) under project DE1915/2-1.

\newpage
\bibliographystyle{IEEEtran}
\bibliography{definitions,references}
\clearpage
\newpage

\IEEEtriggeratref{4}



\end{document}